\documentclass[11pt,amsmath,amssymb,floatfix,aps]{revtex4}
\usepackage{graphicx}

\begin{document}

\title{Time is not the problem.}
\date{\today}
    \author{Olaf Dreyer}
    \email{odreyer@mit.edu}
    \affiliation{Center for Theoretical Physics, Massachusetts Institute of Technology, 
    \\77 Massachusetts Ave, Cambridge, MA 02139}

\begin{abstract} Attempts to quantize general relativity encounter an odd problem. The Hamiltonian that normally generates time evolution vanishes in the case of general relativity as a result of diffeomorphism invariance. The theory seems to be saying that time does not exist. The most obvious feature of our world, namely that time seems to progress and that the world changes accordingly becomes a problem in this presumably fundamental theory. This is called the problem of time. In this essay we argue that this problem is the result of an unphysical idealization. We are caught in this ``problem of time'' trap because we took a wrong turn in the early days of relativity by permanently including a split of geometry and matter into our physical theories. We show that another possibility exists that circumvents the problem of time and also sheds new light on other problems like the cosmological constant problem and the horizon problem in early universe cosmology.
\end{abstract}

\maketitle
\subsection{The problem of time}\label{sec:intro}
In Newtonian physics the nature of time is straightforward. Time has the structure of the real line and is one of the a priori features of the theory. Given the positions and the velocities of particles at one time $t_0$ together with the physical laws we can find the positions and velocities of these particles for all times $t$. This ability to predict the future state of a system given its initial state is the hallmark of physical law. For its formulation we require a notion of time. Physical laws and time go hand in hand.

This simple picture of time was taken for granted until the middle of the last century when the canonical structure of general relativity was understood. Because general relativity is diffeomorphism invariant and because time evolution is nothing but a special type of diffeomorphism the Hamiltonian of the classical theory is nothing but a constraint. 
\begin{equation}\label{eqn:constarint}
H( q, p ) = 0.
\end{equation}
While this is somewhat baffling in the classical theory it becomes an almost intractable problem in the quantum theory. Here the Hamiltonian $H$ becomes an operator that is required to vanish on physical states:
\begin{equation}
H \vert \psi_{\mbox{\scriptsize physical}} \rangle = 0
\end{equation}
The reason why this equation is harder to deal with than equation (\ref{eqn:constarint}) is that presumably $\vert\psi_{\mbox{\scriptsize physical}}\rangle$ represents the quantum mechanical superposition of states corresponding to different classical spacetimes. The question of the classical limit now becomes very hard to deal with. How is one classical spacetime like Minkowski space to emerge from such a superposition?

This set of questions is usually referred to as the problem of time. It has been a central problem in quantum gravity for the last fifty years and remains so to this day. Instead of reviewing the current state of the discussion on the problem of time we want to point out that the problem of time as described above arises because general relativity allows us to tread the gravitational field in isolation. In general relativity the metric $g_{\mu\nu}$ is a new field whose interplay with the other matter fields is described by Einstein's equations. It makes perfect sense to look at the metric alone and it is in this context that the above discussion applies. The problem of time arises because the metric is distinct from matter fields. In this essay we argue that it is this split between geometry and matter that is to blame for the problem of time together with a number of other problems.

\subsection{The split}\label{sec:thesplit}
The split between geometry and matter originated in the early days of special relativity. Before Einstein axiomatized special relativity Lorentz derived many of the important formulae of special relativity in a completely different way. He looked at Maxwell theory and asked himself the question: Given that matter is described by Maxwell's equations what does that imply for our ability to measure space and time intervals? 

Although Lorentz was working in a Newtonian framework he found that Newton's absolute space is veiled from us by the fact that our observations rely on matter that obeys Maxwell's equations. We can see why this is the case by looking at the field sue to a moving charge: 
\begin{eqnarray}
E_x & = & e x (x^2 + y^2 + {z^\prime}^2)^{-3/2}\left(1 - \frac{v^2}{c^2}\right)^{-1/2} \\
E_y & = & e y (x^2 + y^2 + {z^\prime}^2)^{-3/2}\left(1 - \frac{v^2}{c^2}\right)^{-1/2} \\
E_z & = & e z^\prime (x^2 + y^2 + {z^\prime}^2)^{-3/2}\\
B_x & = & -\frac{v}{c} E_y\\
B_y & = & \frac{v}{c} E_x
\end{eqnarray}
with 
\begin{equation}
z^\prime = (z - vt)\left(1 - \frac{v2}{c^2}\right)^{-1/2}
\end{equation}
In the case where $v=0$ this field is spherically symmetric. For $v\ne 0$ the field becomes squeezed by the factor 
\begin{equation}\label{eqn:squeeze}
\sqrt{1 - \frac{v^2}{c^2}}
\end{equation}
in the direction of motion.

Given this behavior of the electromagnetic field it is not hard to see that a solid object made up of atoms whose charged nuclei produce just this kind of field when in motion gets squeezed by the same factor (\ref{eqn:squeeze}). The electrons that surround the nucleus and that are responsible for the chemical bonds holding the object together will change their orbits with the changing field. Their orbits will be squeezed by the same factor (\ref{eqn:squeeze}) and so will the solid object as a whole. Since our measuring devices are such solid objects we find Minkowski space instead of Newton's absolute space. 

Let us now contrast this point of view with the way we currently understand special relativity. Our current understanding originated with Einstein and Minkowski and their axiomatization of special relativity. We think of special relativity as providing us with a background $\eta_{\mu\nu}$. Matter then propagates \emph{on} this background and we demand that is does so inside the null cones of the metric $\eta_{\mu\nu}$. 

These two views of special relativity are very different in how matter relates to geometry. In Einstein's special relativity matter lives on geometry and has to conform to it. In Lorentz's view matter is used to define the geometry. We see that in Einstein's view matter and geometry completely split whereas in Lorentz's view geometry depends on there being matter. 

Einstein's split between geometry and matter carried over into general relativity. The equivalence principle implies that we can always describe the physics locally as if we were in Minkowski space. Globally though we can not glue these spaces together to give just one Minkowski space and we end up with a curved spacetime instead. What general relativity has in common with special relativity though is the fundamental split between geometry and matter.

\subsection{Internal relativity}
This raises the question whether there exists a Lorentz type version of general relativity that does without this split? Internal relativity \cite{dreyerir} is our attempt at constructing such a theory. 

As in the case of special relativity spacetime notions arise from the matter degrees of freedom of the theory. The metric plays no part in the fundamental formulation of the theory. In particular we will not try to quantize the metric. Instead we will work with a many body quantum theory from the start and will construct the metric from the emergent matter degrees of freedom. To get an idea of the kind of system that we have in mind we look at one example: the one-dimensional XY model. It consist of a large number $N$ of spins whose behavior is governed by the Hamiltonian
\begin{equation}\label{eqn:hamil}
H = \sum_{i=1}^N (S^+_iS^-_{i+1} + S^-_iS^+_{i+1}).
\end{equation}
The operators $S^\pm$ are related to the Pauli matrices $\sigma^i$, $i=1,2,3$, by
\begin{equation}
S^\pm = \frac{1}{2}(\sigma^1 \pm \sigma^2).
\end{equation}
This system is so simple that it can be solved completely. Through a Jordan-Wigner transform this Hamiltonian can be written as a sum of free fermions $f$. 
\begin{equation}
H = \sum_{k=1}^N\varepsilon(k) f_k^\dagger f_k,
\end{equation}
with the dispersion relation
\begin{equation}
\varepsilon(k) =  4\pi \cos \frac{2\pi}{N} k. 
\end{equation}
If we fill the sea of fermions with negative energy then the excitations above the sea have the approximately linear dispersion relation
\begin{equation}\label{eqn:linear}
\epsilon(\Delta k) = \frac{8\pi^2}{N}\Delta k.
\end{equation}
These excitations are the elementary particles of the low energy theory. They play the same role here that light plays in special relativity. The linear dispersion relation (\ref{eqn:linear}) gives the light cones of the theory.

This model is unfortunately too simple to go through the kind of reasoning that gave us Minkowski space in the last section. What we can see already though is that the light cone structure of equation (\ref{eqn:linear}) changes the causality from Newtonian to Lorentzian. Fortunately there are more elaborate spin models (see for example  \cite{wen}) of this kind whose low energy physics contains both fermions and gauge bosons and whose dynamics is described by quantum electrodynamics. In these models the arguments of the last section can be carried through directly. 

So special relativity does appear in the low energy physics of non-relativistic spin models. What about general relativity? Can we also get a curved spacetime?

The reason we also get a curved spacetime is that the presence of excitations changes the vacuum around them. Let us look at a simple model again where the ground state is described by an order parameter $\theta$. An excitation in this model is a deviation of the order parameter from its vacuum value $\theta_0$. If we now imagine a bound object \textbf{\textsf{C}} made up of these excitations then in the vicinity of this object the value of the order parameter will be a value $\theta_{\mbox{\scriptsize \textbf{\textsf{C}}}}$ different from the vacuum value $\theta_0$. To find the value of the order parameter in the presence of such a bound object we have to solve the equation
\begin{equation}
\Delta \theta = 0,
\end{equation}
with the boundary conditions
\begin{eqnarray}
\theta & \xrightarrow[\ \ r\rightarrow\infty\ \ ]{} & \theta_0 \\
&\theta \vert_{\partial\mbox{\scriptsize \textbf{\textsf{C}}}} =  \theta_{\mbox{\scriptsize \textbf{\textsf{C}}}} & 
\end{eqnarray}
We see from these equations that the presence of one bound object \textbf{\textsf{C}} has an influence beyond its boundary $\partial$\textbf{\textsf{C}}. In the presence of two bound objects \textbf{\textsf{C}}$_1$ and \textbf{\textsf{C}}$_2$ this leads to a force $F$ between the objects given by \cite{dreyergravity}
\begin{equation}
F \propto \frac{m_1 \cdot m_2}{r^2},
\end{equation}
where
\begin{equation}\label{eqn:mass}
m_i \propto \int_{\partial\mbox{\scriptsize \textbf{\textsf{C}}}_i}\nabla \theta\; d\sigma.
\end{equation}
We see that the order parameter here plays the role of a Newtonian potential. Because the bound objects attract each other the geometry that would be measured by internal observers ceases to be flat and is instead given by
\begin{equation}\label{eqn:metric}
ds^2 = (1 + \kappa\theta) dt^2 - (1 - \kappa\theta)\delta_{ij} dx^i dx^j,
\end{equation}
where $\kappa$ is some irrelevant constant of proportionality. In equation (\ref{eqn:mass}) we have given a formula for the gravitational mass of a bound object. What we have not yet been able to show is that this expression also gives the inertial mass of the object. If we could show this then we would have shown that the equivalence principle holds. Because both geometry and matter are emergent in internal relativity there is no room to also require the equivalence principle as an additional fundamental principle. It has to come to come out form the theory itself. This is the central conjecture in internal relativity:

\begin{description}
\item[Conjecture] If the notions of distance, time, and mass are defined completely internally the equivalence principle follows.  
\end{description}

In the above argument there is already a hint of why this conjecture might be true. If one starts with massless excitations then we need to form bound objects if we want to have mass. But we have just seen that a bound object has a gravitational mass. In this way we see that an inertial mass implies a gravitational mass. What we do not know yet is whether they are equal. 

\subsection{Another look at the problem of time}
Assuming that we take this last hurdle and show that we are really dealing with gravity by showing that the equivalence principle holds we can now state very clearly how we ended up with the problem of time. 

The fundamental theory is a many body quantum system that is governed by a simple Hamiltonian and has no problem of time. It also has no notion of geometry. The geometry that we care about only arises at the level of the emergent matter. Once we have these emergent matter degrees of freedom we can use them to define our geometry. As we have seen above the presence of matter influences the geometry and vice versa. Although we have not completely shown this yet the low energy physics is arguably described by general relativity where matter is viewed as sitting on geometry and geometry reacts to the presence of matter. The problem of time now arises when we forget about the fundamental theory and the intimate connection between geometry and matter and look at geometry in isolation. From the standpoint of internal relativity we see that this is an unphysical idealization and the prize we pay for it is the problem of time. 

\subsection{Other problems}
Apart from the problem of time internal relativity throws new light on other problems in physics too. We will briefly look at the cosmological constant problem and the horizon problem in cosmology.

One of the biggest problems that comes with the conventional split of geometry and matter is the cosmological constant problem. It arises because we have to take all contributions of the matter fields to the energy momentum tensor into account, including the zero point energies of $1/2 \hbar \omega$ for every mode of the field. Summing up all these modes gives
\begin{equation}
\int^{\omega_{\mbox{\scriptsize max}}}d\omega\omega^2\; \frac{1}{2} \hbar\omega \propto\omega_{\mbox{\scriptsize max}}^4.
\end{equation}
Here $\omega_{\mbox{\scriptsize max}}$ is the cutoff frequency. A natural choice for $\omega_{\mbox{\scriptsize max}}$would be the Planck frequency $\omega_{\mbox{\scriptsize Planck}}$. This choice gives a contribution to the energy momentum tensor that is more than one hundred orders of magnitude larger than what we observe today. The weight of the vacuum would be so great that we could not possibly see a spacetime that is roughly flat. 

From the point of view of internal relativity it is easy to see where the above argument goes wrong. Because we treat matter as residing on spacetime we are forced to include the vacuum fluctuations in our calculation of the energy momentum tensor. In internal relativity on the other hand matter does not reside on spacetime but it defines spacetime. The excitations of the underlying theory whose low energy behavior is described by the quantum field theory define the notions of distance, time, and causality. In the language of the quantum field theory we can say that spacetime only appears at the level of one-particle states. In internal relativity there are no zero point energies that contribute to the energy momentum tensor and hence the cosmological constant problem does not arise in the form stated above. What still remains to be shown is why it has the small value that it has (see also \cite{dreyerconstant}). 

In internal relativity spacetime is given by the excitations of the underlying theory. Since different phases of the underlying theory have different sets of excitations these phases  will also give different spacetimes. It is in particular interesting to contemplate what happens in a phase transition. Imagine the system undergoes a phase transition from phase A to phase B. The excitations in phase B correspond to the elementary particles that we see around us. How does the phase transition look from the point of view of an observer in phase B?

We want to point out that such a question can not be meaningfully asked in the usual setup of matter on spacetime. In that setup the notion of a  spacetime is required to even formulate the theory. The matter part of the theory requires a manifold together with a metric for its formulation. The possibility of a creation of spacetime is thus excluded because the setup does not allow it. 

The question of how such a phase transition looks from inside the emergent phase might seem rather esoteric until one realizes that there is a problem in cosmology that arises precisely because of a seeming problem with our causal structure. When we look at the cosmic microwave background we see a very homogenous distribution at roughly three degrees Kelvin. The deviations from homogeneity are of the order of $10^{-5}$. Usually when we encounter such regularity we do not loose much sleep over it because we assume that all this regularity points to is that the system has thermalized. Thermalization does of course require that all the parts of the system have been in causal contact. For the early universe this is where the problem arises. It turns out that in the standard big bang cosmology large parts of the sky never were in causal contact. The homogeneity of the microwave background then becomes an enormous problem. What process made it so smooth? This problem is called the horizon problem. 

The horizon problem was one of the motivations for the theory of inflation. By introducing a phase of exponential grows one can account for the smoothness of the cosmic microwave background while maintaining our current notions of causality. Another possibility is given by internal relativity. What if the horizon problem really indicates that our notion of causality was not always valid? As we have seen above in internal relativity a notion of causality can emerge in a phase transition and not be valid beyond it. If this is the case then the horizon problem immediately goes away because we can not use our notion of causality to draw conclusions about the physics before the phase transition. 

With the advent of high precision microwave background observations cosmology has become very constrained by observational data. Any theory of the early universe has to not only explain the horizon problem but also the exact spectrum of the microwave background radiation. Inflationary cosmology passes this test with flying colors. In \cite{dreyercosmo} we have argued that internal relativity can account for a flat spectrum.  We have furthermore argued that the deviation of the spectrum from flatness might be connected to one of the critical exponents describing the phase transition.

\subsection{Discussion}
The nature of time that we are proposing here looks like a throwback to pre-relativity days. In internal relativity there is a background time that is very much like the time in Newtonian physics. Also, if one overlooks the fact that our models use discrete lattices instead of a smooth manifold, we even have a preferred space. How can this be reconciled with what we know about special and general relativity? It can be reconciled by realizing that the spacetime we should care about is not given by this Newtonian background structure. Rather, it is the structure that is seen by the internal observers of the system that is important. Lorentz has shown that this internal perspective gives the predictions of special relativity without Minkowski space as a background. What we have argued is that we can take this attitude further and also include general relativity. This is the goal of internal relativity.

What then is the nature of time that we are proposing? In internal relativity time appears in two different ways. The first notion of time would be completely familiar to a Newtonian physicist. This time is the background time of the theory. The second notion of time would be less familiar to a Newtonian physicist.  It arises because the presence of a background time does not imply that an observer has access to it. Rather, the behavior of matter determines what the observers will measure. As we have seen this means that the internal time as it is measured by an observers differs from the background time.

Because observers have to use matter to perform their spacetime measurements there is a direct link between matter and geometry. Without matter there is no geometry. We have argued that overlooking this fundamental connection is the source of a number of problems. In this essay we have touched on three such problems: the problem of time, the cosmological constant problem, and the horizon problem in early universe cosmology. 

The problem of time arises because in general relativity, as in special relativity, geometry appears as a distinct stage on which matter propagates. The novelty in general relativity is that geometry now reacts to the presence of matter. We have seen that this split between matter and geometry leads to the problem of time because it allows one to talk about the evolution of geometry in the absence of matter. For general relativity this leads to the Hamiltonian being a pure constraint and thus to the problem of time. If on the other hand one takes the fundamental connection between matter and geometry into account one finds that this problem is the consequence of an unphysical idealization: namely that of geometry without matter. So, time is not the problem, the proper understanding of the origin of geometry is. 

\begin{acknowledgments} I would like to thank F.~Markopoulou for fruitful discussions. Also I would like to thank FQXi for funding this research and the W. M. Keck Foundation Center for Extreme Quantum Information Theory for hosting me at the Massachusetts Institute of Technology. 
\end{acknowledgments}

\end{document}